\documentclass[a4paper,fleqn,usenatbib]{mnras}

\usepackage{newtxtext,newtxmath}

\usepackage[T1]{fontenc}
\usepackage{ae,aecompl}

\usepackage{graphicx}	
\usepackage{amsmath}	
\usepackage{amssymb}	

\usepackage{hyperref}
\title[Bayesian source discrimination]{Bayesian Source Discrimination in Radio Interferometry}
\author[Hague et al.]{
Hague, P. R.,
Ye, H.,
Nikolic, B.,
Gull, S. F.
\\
Astrophysics Group, Cavendish Laboratory, University of Cambridge \\
}

\pubyear{2018}

\begin{document}
\label{firstpage}
\pagerange{\pageref{firstpage}--\pageref{lastpage}}
\maketitle

\begin{abstract}
Methods currently in use for locating and characterising sources in radio interferometry maps are designed for processing images, and require interferometric maps to be preprocessed so as to resemble conventional images. We demonstrate a Bayesian code - \texttt{BaSC} - that takes into account the interferometric visibility data despite working with more computationally manageable image domain data products. This method is better able to discriminate nearby sources than the commonly used \texttt{SExtractor}, and has potential even in more complicated cases. We also demonstrate the correctness of the Bayesian resolving formula for simulated data, and its implications for source discrimination at distances below the full width half maximum of the restoring beam.
\end{abstract}

\begin{keywords}
techniques: interferometric -- techniques: image processing -- methods: statistical -- methods: observational -- methods: data analysis
\end{keywords}

\section{Introduction}

Interferometers produce visibility data rather than images. The process that converts these visibilities to brightness maps means that they differ from images in ways which are relevant in source finding. Incomplete coverage of the $uv$ plane leads to more complex and difficult point spread functions than are assumed by source finding algorithms designed to deal with images.

The distinction between maps and images is made here to underline that the former are more processed data products. In an image each pixel corresponds to a measurement taken by a single detector element. The cells of a map produced through synthesis imaging are products of an inverse Fourier transform and thus have values depending on the entire instrument. Any analysis of these maps is really an analysis of the visibilities via the intermediate data product, even if this is not acknowledged. Therefore such analysis should be done with an understanding of the provenance of the map, in order to produce correct inferences.

The problem of extracting point sources from interferometric maps has been covered mathematically in \cite{tan1987}. Another approach to the problem of source finding in radio interferometry by \cite{lochner2015} applied MCMC directly to visibilities, in order to explore degeneracies between science parameters and instrumental errors, but at a much greater computational cost than dealing with dirty maps. In \cite{hobson2003} a Bayesian approach was shown to be valuable in detecting sources in a noisy image.

The correct identification and characterisation of sources in images - and maps - is an important and continuing problem in astronomy. At present it is standard practice to apply tools designed for use on optical images (such as the \texttt{SExtractor} \cite{bertin1996}) to interferometric maps that have been through the CLEAN algorithm,\citep{hogbom1974,schwarz1978} so as to resemble images.

CLEAN produces inaccurate models of the sky, and based on these models inserts a gaussian restoring beam at every point suspected to be a source. A method of finding sources that uses only the resulting maps is limited in its resolution by the restoring beam. The comparative inability of CLEAN-based methods to discriminate sources shows that the raw instrument data are not being exploited optimally, and that there is scope for improvement of algorithms.

Work such as \cite{trakhtenbrot2017} on the rotation of distant galaxies, and \cite{salak2016} mapping the central kiloparsec of NGC 1808, make inference from extended sources at scales comparable to the size of the CLEAN restoring beam. As one step toward the correct analysis of extended sources, it is necessary to discriminate between points on such scales.

We have developed a Markov Chain Monte Carlo (MCMC) process for the detection and characterisation of sources in dirty maps - \texttt{BaSC}\footnote{\url{https://github.com/petehague/BASC}}, which we have released as a public Python library. This code aims to produce more accurate catalog results from sky maps and overcome the problem of resolving structures smaller than CLEAN resolving beam, while saving the effort of making CLEANed maps.
The method involves marginalisation over a parameter space defined by possible locations and fluxes of sources, so as to compute likelihoods for those sources. The resulting likelihoods can then be fed into a Bayesian statistical model.

In this paper we demonstrate the power of this approach by using simulated dirty maps. We first constrain the positions of point sources, thereby demonstrating the resolving power of the Bayesian technique. Secondly, we show the ability to discriminate between two nearby point sources. This latter test is intended to show the improved accuracy available for feature extraction in extended sources. We compare the performance of \texttt{BaSC} with \texttt{SExtractor} in both of these cases. 

\section{Markov Chain Monte Carlo}

MCMC is a method of finding the likelihood distribution of a parameter space via a random walk. The algorithm produces a chain of samples $\mathbf{X}_{\rm 0...k}$ from the parameter space, such that

\begin{equation}
\mathbf{X}_{\rm n+1} =
\begin{cases}
\mathbf{X}_{\rm n} &  \text{if }P(\mathbf{D} | \mathbf{X}') < P(\mathbf{D} | \mathbf{X}) \\
 & \text{ and } rand > {P(\mathbf{D} | \mathbf{X}') \over P(\mathbf{D} | \mathbf{X})} \\
\mathbf{X}' & \text{otherwise}
\end{cases}
\end{equation}

where $\mathbf{D}$ is the data being considered, $rand$ is a random number between 0 and 1, and $\mathbf{X}'$ is a proposed new sample, which due to the Markov chain principle will be a random function of the previous sample. If new samples are selected from a symmetric function, so that the likelihood of selecting a sample $\mathbf{X}_{\rm a}$ given a current sample $\mathbf{X}_{\rm b}$ is the same as that of selecting $\mathbf{X}_{\rm b}$ given a current sample $\mathbf{X}_{\rm a}$, then the chain will always converge towards a state in which the density of samples in a given volume of parameter space is proportional to the likelihood. Typically, samples are selected from a Gaussian function centred on the current sample; this is the Metropolis-Hastings algorithm \cite{metropolis1953, hastings1970}.

The MCMC driver used here is \texttt{BayeSys},\footnote{\url{https://www.mrao.cam.ac.uk/~steve/algor2008/images/BayeSys_manual.pdf}} developed by \cite{bayesys}. This program differs from other MCMC algorithms through its use of \textit{atoms}. Instead of each model being a set of parameters $\{x_{\rm i}\}$, the model consists of an atom count $n$ and $n$ copies of the parameters $\{\{x_{\rm i}\}_{\rm n}\}$. (In this case, the parameters are coordinates of cells on the dirty maps, together with a flux $f$.) A conventional implementation of MCMC would be equivalent to using a single atom of this sort throughout the chain. The set of atoms is applied to a map to provide a single likelihood to be used in the MCMC process. Multiple models may be generated for each MCMC step.

The atom count is adjusted dynamically by proposing models with one additional atom or one fewer, at random intervals calculated such as to impose a specific prior on the atom count. In all cases in this paper, we use a Poissionian prior with $\alpha=1$:

\begin{equation}
P(n) = (n!e)^{-1}
\end{equation}

Further details of the operation of \texttt{BayeSys} and the control of the number of atoms are set out in \cite{bayesys}.

At the end of the process, atoms from each model are clustered so that they can be associated with sources, after which source properties are extracted from each cluster of atoms.

\section{Likelihood calculation}

Following Chapter 1 of \cite{tan1987}, we start with a set of $N$ point sources having fluxes $\{F_{\rm j}\}$ at sky positions $\{x_{\rm j}\}$, giving a flux $F(\mathbf{x})$ at position $\mathbf{x}$ as follows:

\begin{equation}
F(\mathbf{x}) = \sum_{\rm j=1}^N  F_{\rm j}(\mathbf{x} - \mathbf{x_{\rm j}})
\end{equation}

By means of Fourier transform, this gives a visibility model at a particular visibility $V_{\rm k}$ of the form

\begin{equation}
M_{\rm k} = \sum_{\rm j=1}^N F_{\rm j} e^{2\pi i \mathbf{u_{\rm k} \cdot x_{\rm j}}}
\end{equation}
where $\mathbf{u_{\rm k}}$ is the wavevector of the visibility. For Gaussian errors on the visibilities, the likelihood of the observed visibilities given the model is

\begin{equation}
P(\{V_{\rm k}\} | \{M_{\rm k}\}) =\prod_{\rm k=1}^N \frac{1}{2 \pi \sigma_{\rm k}^{2}} \exp \bigg ( \frac{-|V_{\rm k} - M_{\rm k}|^2}{2\sigma_{\rm k}^2}\bigg )
\end{equation}

The exponential term is equivalent to ${\rm exp}(-\chi^2/2)$. This calculation for $M$ visibilities would require $M\times N$ calculations and may be prohibitively expensive. Alternatively, we can perform only $N^2$ calculations by using the equivalent expression for $\chi^2$,

\begin{equation}
\chi^2 = \sum_{\rm k} \frac{|V_{\rm k}|^2}{\sigma_{\rm k}^2} - \frac{1}{\sigma^2}(2\mathbf{D}^{\mathbf{T}}\mathbf{F} - \mathbf{F}^{\mathbf{T}}\mathbf{B}\mathbf{F})
\label{chisqeq}
\end{equation}

where $\{V_{\rm k}\}$ is the set of visibilities sampled, $\mathbf{D}$ denotes the values of the dirty map at those points, and $\mathbf{B}$ is the corresponding value of the dirty beam,

\begin{equation}
B(\mathbf{x}) = \Re \left[ \sum_{\rm k} {1 \over \sigma_{\rm k}^2} e^{-2\pi i \mathbf{u_{\rm k} \cdot x}} \right] \sigma^2
\label{beameq}
\end{equation}

Since the MCMC chain is constructed entirely from the ratio of log likelihoods $L_{\rm n} - L_{\rm n-1}$, where $L_{\rm n}={\rm constant}-\chi^2/2$,  the first term of Equation \ref{chisqeq} always cancels and does not need to be calculated in order to generate the chain. The calculation need be performed only at the spatial positions of the proposed points. To make use of this likelihood a thorough search of the parameter space is needed, which MCMC is capable of.

In \texttt{BayeSys}, parameters are always in the range $(0,1)$. To meet this condition the parameters $x$ and $y$ are simply given a flat prior and scaled to the size of the map. This works for the small maps being considered here, but will need to be revisited in the case of a large field of view. For the flux parameter, a prior is used such that

\begin{equation}
F = A {f \over 1-f}
\end{equation}
where $f$ is the flux parameter and $F$ is the actual flux of the atom. Here, $A$ is a scaling constant which places the centre of the parameter range ($f=0.5$) at $F=A$. The value of $A$ can be varied to ensure the flux prior encloses the full range of fluxes; setting it equal to the dirty map noise usually gives good results, and this is used as the default.

\texttt{BayeSys} is allowed to burn in for as long as is specified by the annealing alogrithm outlined in \cite{bayesys} (sample from $L^{\lambda}$ with lambda increasing from 0 to 1 with a rate inverseley proportional to the spread of log likelihood values). Then an ensemble of 20 samples per step is generated for the same number of steps as it took to burn in. In this application this typically leads to less than 200 models included in the final output, which is sufficiently accurate here -- although a much larger number of models will have been explored internally by \texttt{BayeSys} without being included. The atoms are then clustered using the DBSCAN algorithm (see Appendix {\ref{clustering}), in order to identify them with sources.

\section{Test Data}

\begin{table}
\centering
\begin{tabular}{| c | c | c | c | c |}
\hline
Set & Images & Points & S/N range & Instrument \\
\hline
A & 5 & 1 & 10-160 & ALMA \\
B & 100 & 2 & 10 & ALMA \\
C & 100 & 2 & 100 & ALMA \\
D & 100 & 2 & 200 & ALMA \\
E & 100 & 2 & 300 & ALMA \\
F & 100 & 2 & 400 & ALMA \\
G & 100 & 2 & 10 & VLA \\
\hline
\end{tabular}
\caption{Properties of the testing sets used in this paper}
\label{tab:settable}
\end{table}

The testing sets are summarised in Table~\ref{tab:settable}. Set A consists of single point source of various signal-to-noise ratios to demonstrate Bayesian resolution. The remaining sets consist of 100 maps containing pairs of points in which one point is at the centre of the map and the other is offset by a random distance and angle. In some cases the offset is zero, giving a single point of double brightness. Set B is the baseline for these sets, with a signal-to-noise ratio of 10. In sets C to F the central point brightness is increased by various factors, but with the distance and angle to the second point unchanged. The standard brightness for these sets is $10\sigma$, where the map noise $\sigma = 5.6\times10^{-4}$Jy/beam. Set G uses simulated VLA maps in order to have a more challenging PSF, reusing the angles and distances.

We use the \texttt{simalma} task in \texttt{CASA} to generate simulated maps. Maps are simulated by inputting a surface brightness map, which in this case is zero everywhere except for cells containing the point sources. This is then converted into an observation using a standard antenna configuration, which also yields both a dirty beam and a primary beam. We use ALMA antenna configuration 2.6 observing for 698s, based on an observation of NGC 1808 as detailed in \cite{salak2016}. This gives a map that is 20.48 arcseconds across, substantially smaller than the primary beam, with 2048x2048 cells of 0.01 arcseconds each.

For set G, the VLA simulation is again performed with CASA, using antenna configuration A, observing for 1 hour, and with the dirty and clean maps having a cell size of 0.03 arcseconds. The flux of the points is 50mJy against a noise of 2mJy. The higher $S/N$ used here compared to the equivalent test for simulated ALMA simulations was determined to give a better performance from \texttt{SExtractor}.

\begin{figure}
\includegraphics[width=\linewidth]{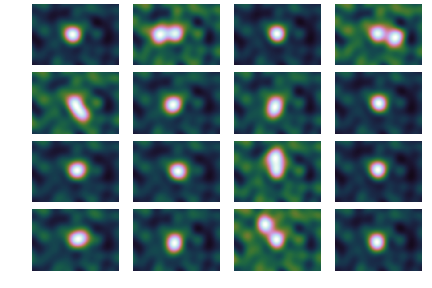}
\caption{Cutouts of the central area of the 16 examples from the set B (indices 68-83, top left to bottom right). Sample number 68 at top left was only detectable as two distinct points for the case in which one point is 40 times brighter than the other. Sample 70 (top row third from left) is a single point. The colour map is CubeHelix \citep{green2011} }
\label{Testset}
\end{figure}

\begin{figure}
\includegraphics[width=0.5\linewidth]{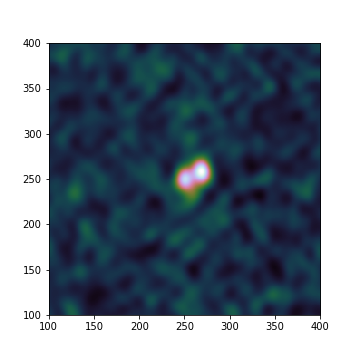}
\includegraphics[width=0.5\linewidth]{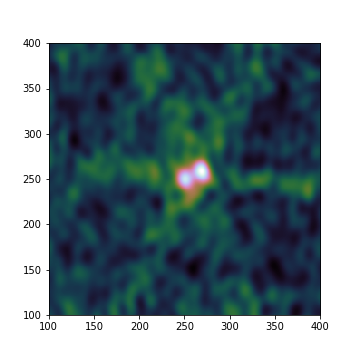}
\caption{Example of a CLEANed (left) and a dirty (right) map from the VLA simulations in set G.}
\label{vlamaps}
\end{figure}

Figure \ref{Testset} shows a sample of the set B, zoomed in on the central region where the two points are located. In each case the points are separated by a uniformly random distance between 0.1 and 1 arcsecond. The angle of the line between the two points is distributed uniformly. The combination of a uniform distance and angle produces a centrally concentrated distribution of points; because the final result will be binned by radius, this distribution is desirable so that each bin has the same number of samples.

We have produced further sets using the same source positions but with the brightness of the source at the centre of the map increased by factors of 10, 20, 30 and 40. We have also used the same set of points to generate simulated VLA observations, so as to determine how a more complex PSF influences the performance of the two methods.

Dirty maps produced from VLA observations are noticeably less clean that those produced from ALMA as a result of differences in their $uv$ coverage, and we therefore expect source detection methods that are dependent on CLEAN to perform less well.

\begin{figure*}
\includegraphics[width=0.45\linewidth]{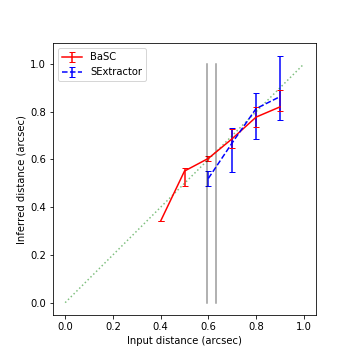}
\includegraphics[width=0.45\linewidth]{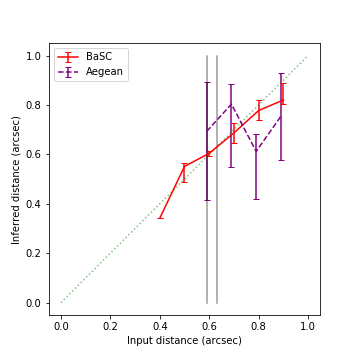}
\caption{Plots of the separation of sources in the input file versus the separation inferred using each source finder using set B. The left shows the comparison with SExtractor and the right shows the comparison with Aegean. The dotted green line represents agreement with the input, solid red shows the results from BaSC, dashed blue shows the results from SExtractor, and dashed purple shows Aegean. Error bars are 90\% confidence intervals. The vertical lines show the sizes of the minor and major axes of the CLEAN restoring beam.}
\label{result2}
\end{figure*}

\begin{figure}
\includegraphics[width=\linewidth]{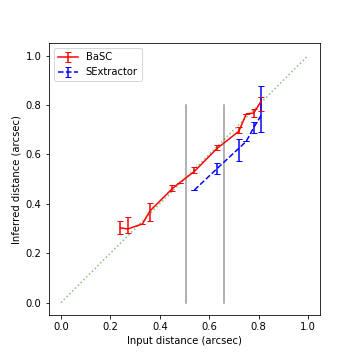}
\caption{Plot of the separation of points in the input file versus the separation inferred using each source finding method, in the case of the VLA simulations (set G) compared against the performance of \texttt{SExtractor}.}
\label{vlafigure}
\end{figure}

\begin{figure*}
\includegraphics[width=0.45\linewidth]{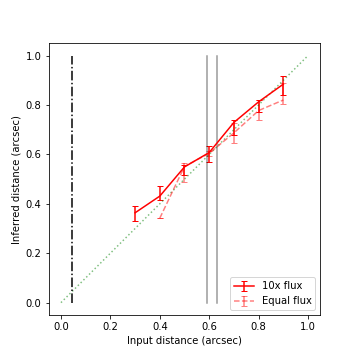}
\includegraphics[width=0.45\linewidth]{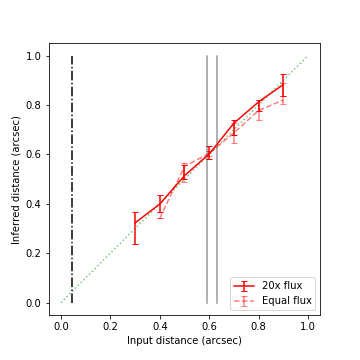}
\includegraphics[width=0.45\linewidth]{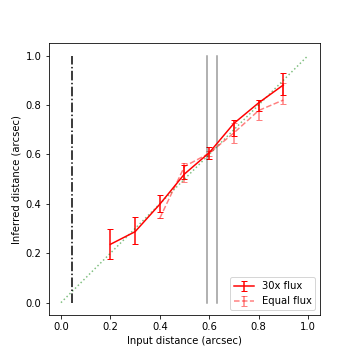}
\includegraphics[width=0.45\linewidth]{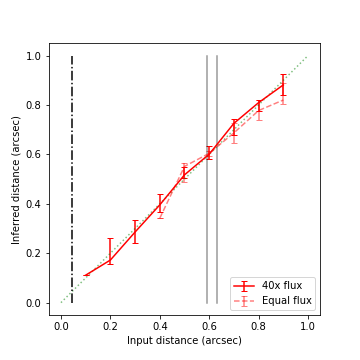}
\caption{Separation of two points when one of the sources is increased in brightness by factors of 10, 20, 30 and 40. The noise and the brightness of the other point are unchanged. The lighter dashed line represents the \texttt{BaSC} performance in the case of equal brightness. The dot-dashed line on the right indicates the detection limit for the dimmer source in the presence of a brighter source (see text).}
\label{resultdiff}
\end{figure*}

\subsection{Bayesian resolving power}
\label{resolving}

Since the maps of atom positions show structure on a substantially smaller scale than the restoring beam or the central peak of the dirty beam, it is necessary to investigate the limit of the resolution using this method.

We first consider the ability of the method to constrain the position of a single source. We begin with the expression for a dirty map $D(\mathbf{x})$ containing a single source of flux $F$ at the origin with visibilities $V_{\rm k},(
\mathbf{u_{\rm k}})$

\begin{equation}
D(\mathbf{x}) \simeq FB(\mathbf{x}) = F \sigma^2 \sum_{\rm k} {1 \over \sigma_{\rm k}^2} {\rm cos}(2\pi \mathbf{u_{\rm k} \cdot x})
\end{equation}

where $B(\mathbf{x})$ is the dirty beam. Truncation of the Taylor expansion of the expression for the beam (equation \ref{beameq}) at the origin gives a Gaussian approximation of the probability that the point is at position $\mathbf{x}$,

\begin{equation}
P(\mathbf{x}) \propto {\rm exp} \left[ -2\pi^2F^2    \sum_{\rm ijk} {(u_{\rm k})_{\rm i} (u_{\rm k})_{\rm j} \over \sigma_{\rm k}^2} x_{\rm i} x_{\rm j} \right]
\label{probeq}
\end{equation}
which implies a covariance of

\begin{equation}
\label{widtheq}
<\delta x_{\rm i} \delta x_{\rm j}> = {\sigma^2 \over F^2} \Delta x_{\rm ij}^2
\end{equation}
where

\begin{equation}
(\Delta x^2_{\rm ij})^{-1} = - \partial_{\rm ij}^2 B(\mathbf{x})\left.\right|_0 = 4\pi^2\sigma^2\sum_{\rm k}  {(u_{\rm k})_{\rm i} (u_{\rm k})_{\rm j} \over \sigma_{\rm k}^2}
\end{equation}
This expression describes the characteristic width of the central part of the beam, for which a sufficient approximation is provided in terms of an ellipse $a,b,\theta$ by the \texttt{CASA} pipeline. The size of the resolving ellipse is therefore just $a/S,b/S$, where $S$ is the signal-to-noise ratio $F/\sigma$.

\begin{table}
\centering
\begin{tabular}{| c | c | c | c |}
\hline
$S/N$ ratio & $\sigma_{\rm x}$ & $\sigma_{\rm y} $ & Prediction \\
\hline
160 & 0.292 & 0.288 & 0.147 \\
80 & 0.574 & 0.547 & 0.294 \\
40 & 0.670 & 0.634 & 0.587 \\
20 & 1.394 & 1.190 & 1.175 \\
10 & 2.188 & 2.386 & 2.351 \\
\hline
\end{tabular}
\caption{Accuracy of positions in processed maps versus theoretical predictions based on Bayesian resolving power formula. Characteristic width of the beam used in this calculation is taken from the value produced by CASA for use in the CLEAN process, although the maps used here have not been CLEANed.}
\label{sntab}
\end{table}

To test this result, we applied BaSC to test set A, generating a set of proposed models for the location of the point, with signal-to-noise ratios ranging from 10 to 160.

\subsection{Other source finding methods}

We use version 2.5 of \texttt{SExtractor} as a comparison for our tests, with the detection threshold set to $3\sigma$ and the minimum deblending contrast parameter set to $0.001$. This parameter indicates the ratio in the integrated flux between two proposed sources if the same island of detected pixels would be considered separate sources. This is the best performing set of parameters that we were able to find. \texttt{SExtractor} uses CLEANed versions of the set B dirty maps that are being tested with \texttt{BaSC}.

For a broader comparison we also test \texttt{AEGEAN}\footnote{\url{https://github.com/PaulHancock/Aegean}}. We tested this with CLEANed and dirty maps, and with both a fixed noise level and with a noise model derived using the \texttt{BANE} command that is part of the same package. We found that the dirty map with a fixed noise level gave the best result in terms of seperation values, however it did require manual removal of outliers outside the area of the map being studied.

\section{Results}

The result of the test on set A, the uncertainty in the position of a single point as a function of signal-to-noise ratio is shown in Table \ref{sntab}. The standard deviation of the generated models along each axis was a good match to the theoretical prediction, and tended towards $1/\sqrt {12}$ of the cell width at the highest $S/N$ i.e. the expected uncertainty within a single cell.

The clustered output of the MCMC chain yields a distance between the two sources, which can be compared to the input parameter. Figure \ref{result2} shows the results, in comparison with the performance of \texttt{SExtractor}.

Out of 100 samples, \texttt{SExtractor} was able to discriminate two distinct points in 27 cases, with a minimum discrimination distance of 0.6 arcseconds, whereas \texttt{BaSC} was able to find two points in 51 cases, with a minimum discrimination distance of 0.3 arcseconds. The major and minor axes of the \texttt{CLEAN} restoring beam are respectively 0.63 and 0.59 arcseconds.

Figure \ref{result2} also shows the same comparison with \texttt{Aegean}. As its performance also depends on the image being cleaned \citep{hancock2018}, this software can be expected to perform comparably to \texttt{SExtractor} in this specific test. Indeed, \texttt{AEGEAN} no longer detected the second point at the same distance as \texttt{SExtractor}. Outliers at a distance of $>1$ arcsecond had to be removed from the \texttt{Aegean} results.

Figure \ref{resultdiff} shows the effect of brightening one of the sources. The ability to discriminate increases with the brightness of the central source, because the constraint of that source becomes tight and largely independent of the dimmer source. The detection of the dimmer source can then be considered an independent problem, with an additional source of noise. As the signal to noise ratio of the brighter source increases, the removal of the main source becomes much more accurate, leaving the uncertainty in the flux as the only significant source of error. For a dim source close to a sufficiently bright source, the detection problem becomes a single point problem with a map noise equal to $2\sigma$. The dotted line in the Figure \ref{resultdiff} shows the resolving power (as discussed in section \ref{resolving}) for the dim source if the noise is $2\sigma$. This is included as an estimate of the lower limit for discrimination of the sources, in the case of a large difference in brightness, on the basis that inside this distance the $1\sigma$ position error would encompass the other point.

\subsection{VLA simulations}

Figure \ref{vlafigure} shows a comparison between \texttt{BaSC} and \texttt{SExtractor} on set G, the VLA data. As well as \texttt{BaSC} being better able to differentiate the sources, there is a systematic underestimation of the distance between them using \texttt{SExtractor}. This is not a consequence of incorrectly fitting Gaussians to the clean map, but is because CLEAN itself has not represented the targets accurately. In Figure \ref{vlamaps} we see that parts of the dirty beam have been left in the map, creating a spurious `bridge' of flux between the two points which causes \texttt{SExtractor} to fit the Gaussians with closer centres.

It is likely that a user of CLEAN could produce maps that represent the sources more accurately in interactive mode. In contrast, \texttt{BaSC} working on the dirty map produced the result shown with no interactivity; the only human input was an initial estimate of the map noise, and basic parameters for the MCMC chain that remained unchanged in all of the present tests.

\section{Conclusion}

We have shown that \texttt{BaSC} is superior to \texttt{SExtractor} and \texttt{Aegean} at the specific task of discriminating between nearby points in interferometric maps. This is hardly surprising, given that \texttt{SExtractor} is not designed for that task, and in fact relies on CLEAN to do the work of finding sources and then reinserting a Gaussian restoring beam back into the map for \texttt{SExtractor} to find.

The CLEAN process produces maps that are readily understandable to humans, but this is not a necessary nor even desirable property when the map is then to be processed by an algorithm not subject to human perceptual norms. Information can only be lost by the insertion of this step into the pipeline, and given that a well tested pixel-based source finder such as \texttt{SExtractor} was less able to recover distinct points from CLEANed maps suggests that this loss of information is impactful on the scales studied in this paper.

The theoretical resolving power for the Bayesian method has been shown to be realised for a single point by using \texttt{BaSC}. The ability to discriminate two points is improved when a bright source is close to a dim source. This is due to the better constraint and consequently better subtraction of the brighter source, and the ability to discriminate in this case has been confirmed by testing to be close to the theoretical limit.

The generalisation of the atomic MCMC method to cases where atoms represent a more complex object than a point source. This will be one aim of future development of \texttt{BaSC}

\section{Acknowledgements}

This  project  has  received  funding  from  the  European Union's Horizon 2020 research and innovation programme under grant agreement No 653477. We wish to thank the referee Paul Hancock for useful feedback.

\bibliographystyle{mnras}

\bibliography{citations}

\appendix

\section{Clustering}
\label{clustering}

Because of the random nature of the MCMC algorithm, atoms in each model are not necessarily assigned to the same source in every instance. It is therefore necessary to adopt a clustering method in order to study individual sources on the map.

The clustering algorithm must be robust enough to disregard these outliers and provide the number of clusters correctly. In simulations the number of clusters is known, such as 2 in the example set out above. For real observations, this number would not be known in advance. We also wish to apply this method without large-scale human intervention. We must therefore choose a clustering algorithm that does not require the number of clusters to be known in advance.

The density-based Spatial Clustering of Applications with Noise (DBSCAN) algorithm meets these requirements. DBSCAN was proposed by \cite{ester1996}, and has been one of the most widely used clustering algorithm ever since. DBSCAN assigns atoms to a cluster or tags them as outliers, based upon the local density.
\cite{schubert2017} explain how DBSCAN can be adapted to process MCMC output:

\begin{enumerate}
\item For each atom in the MCMC output, find the points which are within a radius $\epsilon$ of it.
\item The data point is marked as a seed if the neighbour count is greater than $min_{\rm samples}$, the minimum grouping number
\item For each seed, find all points connected to it by steps no larger than $\epsilon$. Mark as outliers those unconnected to a seed in this way.
\item Each group of points containing one or more seeds is considered as a cluster, where MCMC considers there to be a peak in the likelihood distribution. The location of the source is calculated as the mean of the location of points within this cluster, while the outliers are removed in order to avoid distortion of the data.
\end{enumerate}

Although DBSCAN does not require the number of clusters and does not constrain their shape, it does need the radius of the cluster searching area $\epsilon$ as well as the minimum grouping number $min_{\rm samples}$. This can be a challenge in practice:
\begin{itemize}
\item The choice of $\epsilon$ should not be too large, or else two sources close to each other would be considered as one; whereas if $\epsilon$ is too small then outliers would be considered as sources, or even single sources would be divided into two or more.
\item The minimum grouping number $min_{\rm samples}$ should not be too small, or else outliers would be considered as sources; whereas if the number is too large then two close-by sources would be considered as one, and in the worst case even the sources would be considered as outliers.
\end{itemize}

\begin{figure}
\includegraphics[width=\linewidth]{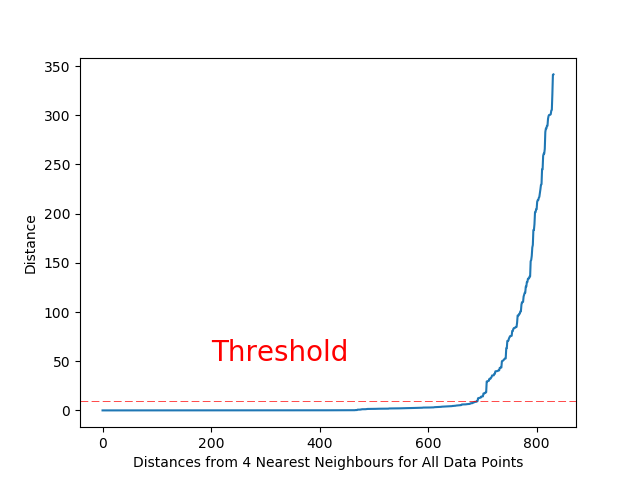}
\caption{The x-axis represents the distances of all data points from their 4 nearest neighbours in ascending order, and the y-axis represents the distances. The sorted k-dist graph is used to provide an estimate of $\epsilon$ by classifying outliers and non-outliers according to a threshold, shown by the broken line. The distance value at the broken line is the recommended value of $\epsilon$}
\label{distanceplot}
\end{figure}

\begin{figure}
\centerline{\includegraphics[width=\linewidth]{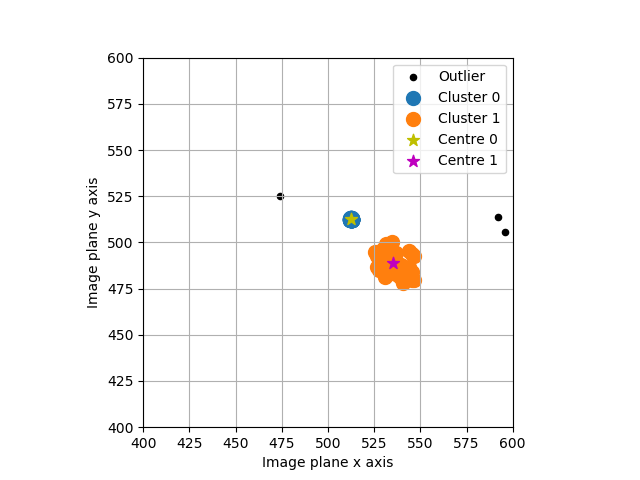}}
\caption{BaSC output of simulated two-source data; one source is 10 times brighter than the other. The black dots on the map are outliers, and the blue and orange dots are two distinct clusters with stars as their cluster centres. The two sources have a separation of 30 cells.}
\label{dbscan01}
\end{figure}

Since $\epsilon$ and $min_{\rm samples}$ work as a pair, in this application of MCMC results the value of $\epsilon$ should not be less than 0.5, or half of the pixel size. This is to prevent atoms in the same pixel being separated into more than one cluster. Also, $min_{\rm samples}$ is not less than 10 so that outliers are not considered as clusters. In using the program there will be a recommended value of $\epsilon$ for each dataset with $min_{\rm samples} = 10$. The recommendation was made based on the ``sorted k-dist graph'' \citep{ester1996}; see Figure \ref{distanceplot}. This figure plots the distances from $K- nearest$ neighbours for all data points. In this case we set $K = 4$; the x-axis represents the distances of all data points from their $4-nearest$ neighbours in ascending order, and the y-axis represents the distance values.  ``Sorted k-dist graph'' is used to give a rough estimate of $\epsilon$ by classifying outliers and non-outliers according to a threshold, shown as a broken line in Figure \ref{distanceplot}. Package users can use the recommended $\epsilon$ directly based on our threshold searching program, or can choose $\epsilon$ intuitively.

\begin{figure}
\centerline{\includegraphics[width=\linewidth]{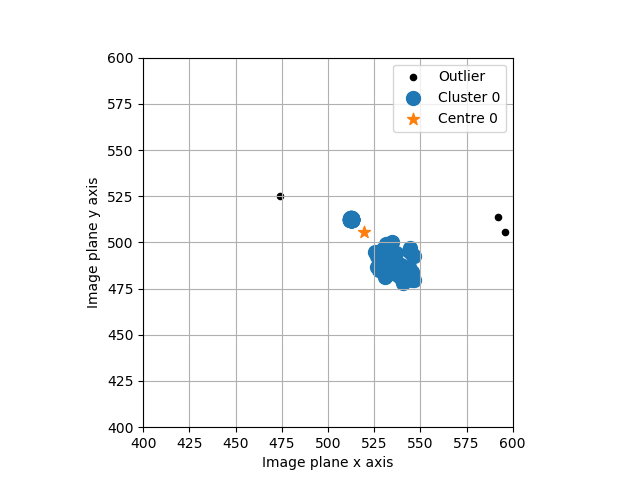}}
\caption{The black dots on the map are outliers, while the blue dots represent the only cluster with a star as its cluster centre. Upon increasing $\epsilon$ to 30, the two clusters detected in Figure \ref{dbscan01} are considered as one.}
\label{Demo_results_98_2}
\end{figure}

\begin{figure}
\includegraphics[width=\linewidth]{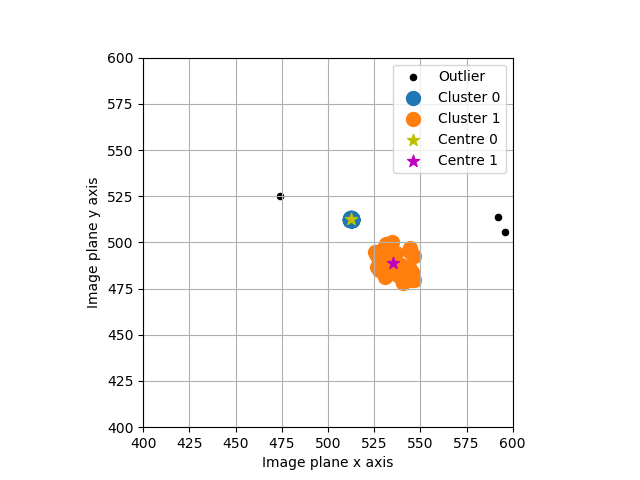}
\caption{The black dots on the map are outliers, while the blue and orange dots are two different clusters, with stars as their respective cluster centres. The two clusters were easily found by HDBSCAN with their centers at (512.50, 512.50) and (535.50,488.50). The two parameters chosen are $min_{\rm cluster}$= 10 and $min_{\rm samples}$= 30}
\label{HDBSCAN_result_9}
\end{figure}

A more recent clustering algorithm, Hierarchical Density-Based Spatial Clustering of Applications with Noise (HDBSCAN), has been developed by \cite{campello2013} based on DBSCAN, and does not need the user to provide $\epsilon$. HDBSCAN is another density-based clustering algorithm, but unlike DBSCAN it can find clusters with varying densities. Only two parameters are required by the hdbscan library \footnote{\url{https://github.com/scikit-learn-contrib/hdbscan}}. The first is $min_{\rm cluster}$, which is identical to $min_{\rm samples}$ in DBSCAN, the minimum grouping number. The second is $min_{\rm samples}$, which controls the sensitivity of HDBSCAN for picking up less dense clusters; for larger $min_{\rm samples}$, the algorithm is more likely to ignore a number of outliers which cluster together; for smaller $min_{\rm samples}$ those outliers may be regarded as clusters. Consequently, HDBSCAN can be used in this application; it is recommended that $min_{\rm cluster} = 10$ at first, in order to pick up as many clusters as possible, then set $min_{\rm samples} = 30$ so that lower density agglomerations, which are more likely to be outliers, are not taken as clusters.

When dealing with a large number of maps without human supervision, we recommend DBSCAN with the recommended $\epsilon$, or HDBSCAN with $min_{\rm cluster} = 10$ and $min_{\rm samples} = 30$. After the clustering process, and if concerns become obvious over certain maps, parameters can be changed manually by the user so  as to improve the clustering process in specific applications.

Figure \ref{dbscan01} shows an example of implementing DBSCAN. The black dots on the map are outliers, and the blue and orange dots are two distinct clusters with amber stars and red stars as their respective cluster centres. The two clusters were readily identified by DBSCAN, with their centres at (512.50, 512.50) and (535.50, 488.50). DBSCAN has an outlier detection feature, but asks for the radius of cluster sizes as well as the minimum grouping number. The recommended radius parameter is 10, and the minimum grouping number was chosen as 10. There are 43 outliers, marked as black dots. If instead $\epsilon = 20$ and $min_{\rm samples}$ = 10, the two clusters are considered to be centred at the single location (519.50, 505.50). This is because the radius $\epsilon$ is too large, so that DBSCAN considers the two cluster centres separated by 33.24 ($<$ 40) units as a single cluster; see Figure \ref{Demo_results_98_2}.

Figure \ref{HDBSCAN_result_9} shows an example of implementing HDBSCAN. The black dots on the map are outliers, and the blue and amber dots are two distinct clusters with stars as their respective cluster centres. The two clusters were readily found by HDBSCAN, with their centers at (512.50, 512.50) and (535.50,488.50). The two parameters chosen are $min_{\rm cluster}$= 10 and $min_{\rm samples}$= 30.

\bsp	
\label{lastpage}
\end{document}